\documentclass[aps,nofootinbib,preprintnumbers]{revtex4}
\usepackage{lipsum}
\usepackage{CJK}
\usepackage{amsfonts}
\usepackage{amsmath}
\usepackage{amssymb}
\usepackage[english]{babel}
\usepackage{graphicx}
\usepackage{epsfig}
\usepackage{bm}
\usepackage{verbatim}
\usepackage[utf8]{inputenc}
\usepackage{booktabs}
\usepackage{multirow}
\usepackage{subfig}
\usepackage{slashed}
\usepackage{xcolor}
\usepackage[colorlinks=true,urlcolor=red,citecolor=red]{hyperref}
\usepackage[font=small]{caption}
\usepackage{float}
\usepackage{blindtext}
\usepackage{placeins}
\usepackage[utf8]{inputenc}
\usepackage{bbm}
\usepackage[colorinlistoftodos]{todonotes}

\newenvironment{Eqnarray}{\arraycolsep 0.14em\begin{eqnarray}}{\end{eqnarray}}
\newcommand{\ba}{\begin{Eqnarray}}
\newcommand{\ea}{\end{Eqnarray}}
\newcommand{\be}{\begin{equation}}
\newcommand{\ee}{\end{equation}}
\newcommand{\bal}{\begin{aligned}}
\newcommand{\eal}{\end{aligned}}
\newcommand{\bea}{\begin{eqnarray}}
\newcommand{\eea}{\end{eqnarray}}
\newcommand{\ben}{\begin{enumerate}}
\newcommand{\een}{\end{enumerate}}
\newcommand{\bit}{\begin{itemize}}
\newcommand{\eit}{\end{itemize}}
\newcommand{\bde}{\begin{widetext}}
\newcommand{\ede}{\end{widetext}}

\def\lsim{\mathrel{\rlap{\lower4pt\hbox{\hskip1pt$\sim$}}
\raise1pt\hbox{$<$}}}
\def\gsim{\mathrel{\rlap{\lower4pt\hbox{\hskip1pt$\sim$}}
\raise1pt\hbox{$>$}}}

\def\3211{$\mathrm{SU(3) \otimes SU(2)_L \otimes U(1)_R \otimes U(1)_{B-L}}$ }
\def\321{$\mathrm{SU(3) \otimes SU(2) \otimes U(1)}$ }
\def\422{$\mathrm{SU(4) \otimes SU(2) \otimes SU(2)_R}$ }

\newcommand{\mathsym}[1]{{}}

\definecolor{bostonuniversityred}{rgb}{0.8, 0.0, 0.0}

\newcommand{\Antonio}[1]{{#1}}

\newcommand{\Yoxara}[1]{{#1}}
\usepackage{soul}
\input{tcilatex}
\begin{document}

\title{An extended 3-3-1 model with radiative linear seesaw mechanism}
\author{A. E. C\'arcamo Hern\'andez$^{a,b,c}$}
\email{antonio.carcamo@usm.cl}
\author{Sergey Kovalenko$^{b,c,d}$}
\email{sergey.kovalenko@unab.cl}
\author{Farinaldo S. Queiroz$^{c,e,f}$}
\email{farinaldo.queiroz@ufrn.br}
\author{Yoxara S. Villamizar$^{e}$}
\email{yoxara@ufrn.edu.br}
\affiliation{$^a$ Universidad T\'{e}cnica Federico Santa Mar\'{\i}a,
Casilla 110-V, Valpara\'{\i}so, Chile,\\
$^{{b}}$ Centro Cient\'{\i}fico-Tecnol\'ogico de Valpara\'{\i}so, Casilla 110-V, Valpara\'{\i}so, Chile\\
$^{c}$ Millennium Institute for Subatomic physics at high energy frontier - SAPHIR, Fernandez Concha 700, Santiago, Chile\\
$^{d}$ Departamento de Ciencias F\'isicas, Universidad Andr\'es Bello,
Sazi\'e 2212, Piso 7, Santiago, Chile\\
$^{e}$ International Institute of Physics, Universidade Federal do Rio
Grande do Norte,
Campus Universitario, Lagoa Nova, Natal-RN 59078-970, Brazil\\
$^{f}$ Departamento de Fisica, Universidade Federal do Rio Grande do Norte,
59078-970, Natal, RN, Brasil\\
}
\date{\today }

\begin{abstract}
Motivated by the recent muon anomalous magnetic moment (g-2) measurement at FERMILAB and non-zero neutrino masses, we propose a model based on the $SU(3)_C  \times SU(3)_L \times U(1)_X$ (3-3-1) gauge symmetry. The most popular 3-3-1 models in the literature require the presence of a scalar sextet to address neutrino masses.  In our work, we show that we can successfully implement an one-loop linear seesaw mechanism with right-handed neutrinos, and vector-like fermions to nicely explain the active neutrino masses, and additionally reproduce the recent Muon g-2 result, in agreement with existing bounds. 
\end{abstract}

\maketitle

\section{Introduction}
\label{intro}
Despite its success in describing the electromagnetic, strong and weak interactions with a high degree of precision, the Standard Model (SM) has several unaddressed issues. The very large mass hierarchy in the SM fermion sector extending, over a range of about 13 orders of magnitude, from the light active neutrino mass scale up to the top quark mass, the small quark mixing angles and the sizeable leptonic mixing ones, the existence of three fermion families, the electric charge quantization as well as the muon anomalous magnetic moments do not find an explanation within the context of the SM. To address these issues, we propose a renormalizable theory based on the $SU\left( 3\right) _{C}\times SU\left( 3\right) _{L}\times U\left( 1\right) _{X}$ gauge symmetry (3-3-1 model), which is suplemmented by the spontaneously broken $U(1)$ global symmetry. In our proposed theory, we considered the $SU(3)_C\times SU(3)_L\times U(1)_X$ gauge symmetry, since models having this symmetry naturally explain the number of SM fermion families as well as the electric charge quantization, see for instance \cite{Valle:1983dk,Pisano:1991ee,Frampton:1992wt,Hoang:1995vq,CarcamoHernandez:2005ka,Chang:2006aa,Hernandez:2013mcf,Hernandez:2013hea,Boucenna:2014ela,Hernandez:2014lpa,Hernandez:2014vta,Okada:2015bxa,Hernandez:2016eod,Fonseca:2016tbn,CarcamoHernandez:2017cwi,CarcamoHernandez:2018iel,CarcamoHernandez:2019vih,CarcamoHernandez:2019iwh,CarcamoHernandez:2019lhv,CarcamoHernandez:2020pxw,Das:2020pai,CarcamoHernandez:2020ehn}. Unlike several 3-3-1 models, our proposed theory does not have right handed Majorana neutrinos in the fermionic spectrum that induce a low scale linear or inverse seesaw mechanism that produces the tiny neutrino masses. In our proposed model, the heavy charged exotic vector like leptons are included instead of the right handed Majorana neutrinos and they induce a low scale Universal seesaw mechanism that produces the SM charged lepton masses. These heavy charged exotic leptons mediate a radiative linear seesaw mechanism that produces the tiny masses of the light active neutrinos and at the same time are crucial for accommodating the muon anomalous magnetic moment. In our proposed model, the third family of the SM quarks receive tree-level masses from their Yukawa interactions with the $\eta$ and $\rho$ $SU(3)_L$ scalar triplets, whereas the first and second families of SM quarks get their masses from a Universal seesaw mechanism mediated by heavy vector like quarks. \Antonio{It is worth mentioning that the Universal Seesaw mechanism have been implemented to generate the masses of all SM charged fermions and light active neutrinos for the first time in \cite{Davidson:1987mh,Davidson:1987mi,Davidson:1989bx}, in the context of left-right symmetric models and $SU(5)$ Grand Unified Theory. Here we implement the Universal Seesaw mechanism only to generate the SM charged lepton masses as well as the masses of the first and second families of SM quarks, in the context of a model with extended gauge symmetry $SU(3)_C\times SU(3)_L\times U(1)_X$ gauge symmetry, which is different than the models considered in \cite{Davidson:1987mh,Davidson:1987mi,Davidson:1989bx}.}

The layout of the reminder of the paper is as follows. In section \ref{model} we describe our proposed 3-3-1 model. Its implications on the SM fermion mass hiearchy are discussed in section \ref{fermionmasses}. The model scalar potential is analyzed in detail in section \ref{scalarpotential}. The implications of our model in the muon anomalous magnetic moment are discussed in section \ref{gminus2}. We conclude in section \ref{conclusions}.

\section{The model}
\label{model}

We propose a renomalizable theory based on the $SU\left( 3\right) _{C}\times
SU\left( 3\right) _{L}\times U\left( 1\right) _{X}$ gauge symmetry (3-3-1
model), which is supplemented by the spontaneously broken $%
U\left( 1\right) $ global symmetry, where the scalar sector is composed of 
three $SU\left( 3\right) _{L}$ scalar triplets and two gauge singlet
scalars, from which one is electrically neutral and the other one
electrically charged. The fermion sector of our theory contains the
fermionic particles of the conventional 3-3-1 model with right handed
neutrinos plus the following $SU\left( 3\right) _{L}$ singlets charged
vector like fermions: two exotic up type quark $T_{n}$, two exotic down type
quarks $B_{n}$ ($n=1,2$)\ and three exotic charged leptons $E_{i}$ ($i=1,2,3$
) in singlet representations under $SU\left( 3\right) _{L}$. The
aforementioned exotic charged fermion spectrum is the minimal necessary so
that no massless charged SM-fermions would appear in the model. Such
extension of the particle content of the 3-3-1 model with right handed
neutrinos allows the implementation of the tree-level Universal seesaw
mechanism to generate the SM charged fermion mass hierarchy as well as of
the one-loop level linear seesaw mechanism to produce the light active
neutrino masses. It is worth mentioning that a successfull implementation of
the tree-level Universal seesaw mechanism to generate the SM charged fermion
mass hierarchy, requires to extend the scalar spectrum of the conventional
3-3-1 model by adding a electrically neutral gauge singlet scalar field $%
\sigma $ which allows mixings between left handed charged exotic fermions
and right handed SM fermions. The mixing between left handed SM charged
fermions and right handed exotic charged fermions arises from renormalizable
Yukawa interactions involving the $\eta $ and $\rho $ $SU(3)_{L}$ scalar
triplets. Furthermore, the implementation of the one loop level radiative
linear seesaw mechanism to generate the light active neutrino masses
requires to consider the electrically charged scalar singlet $S^{-}$ in the
scalar spectrum. The scalar, quark and lepton content with their assignments
under the $SU(3)_{C}\times SU(3)_{L}\times U(1)_{X}\times U\left( 1\right) $
group are shown in Tables \ref{scalars}, \ref{quarks} and \ref{leptons},
respectively. In our proposed model the top and bottom quarks as well as the
exotic charged fermion masses are generated at tree level. The top and
bottom quarks get their masses from renormalizable Yukawa interactions
involving different $SU\left( 3\right) _{L}$ scalar triplets ($\eta $ and $%
\rho $ for the top and bottom quarks, respectively) thus allowing an
explanation of their mass hierarchy, which is consequence of the VEV
hierarchy of the neutral components of these triplets. The $U\left( 1\right) 
$ global symmetry prevents the appearance of renormalizable Yukawa
interactions that can directly generate tree level masses for the first and
second families of SM quarks as well as for the SM charged leptons. The
masses of the first and second families of SM quarks as well as the SM
charged lepton masses are generated from a tree level Universal seesaw
mechanism mediated by heavy charged vector like fermions. In addition, light
active neutrino masses are generated from a one loop level radiative linear
seesaw mechanism mediated by charged exotic vector-like leptons and
electrically charged scalars.

In addition, our model does not have fermions with exotic electric charges.
Thus, in our model the electric charge is defined as:

\begin{equation*}
Q=T_{3}-\frac{1}{\sqrt{3}}T_{8}+X,
\end{equation*}

In our model the full symmetry $\mathcal{G}$ experiences the following
spontaneous symmetry breaking chain: 
\begin{eqnarray}
&&\mathcal{G}=SU(3)_{C}\times SU(3)_{L}\times U(1)_{X}\times U(1){\ %
\xrightarrow{v_{\chi}}}  \notag \\
&&\hspace{7mm}SU(3)_{C}\times SU(2)_{L}\times U(1){\xrightarrow{v_\eta
,v_{\rho},v_{\sigma}}}  \notag \\
&&\hspace{7mm}SU(3)_{C}\times U(1)_{Q},  \label{SB}
\end{eqnarray}
where the different symmetry breaking scales fulfill the hierarchy: 
\begin{equation}
v_{\chi }\gg v_{\eta },v_{\rho }\sim v_{\sigma },  \label{VEVhierarchy}
\end{equation}
with $v_{\eta }^{2}+v_{\rho }^{2}=v^{2}$, $v=246$ GeV. The first step of
spontaneous symmetry breaking is produced the $SU(3)_{L}$ scalar triplet $%
\chi $, whose third component acquires a $10$ TeV scale vacuum expectation
value (VEV) that breaks the $SU(3)_{L}\times U(1)_{X}$ gauge symmetry, thus
providing masses for the exotic fermions, non Standard Model gauge bosons
and the heavy CP even neutral scalar state of $\chi $. The second step of
symmetry breaking is caused by the gauge singlet scalar $\sigma $ as well as
by the remaining two $SU(3)_{L}$ scalar triplets $\eta $ and $\rho $, whose
first and second components, respectively, get VEVs at the Fermi scale, thus
producing the masses for the SM particles and for the physical neutral
scalar states arising from those scalar triplets. We assume that the scale $%
v_{\chi }$ of spontaneous $SU(3)_{L}\times U(1)_{X}$ gauge symmetry breaking
is about $10$ TeV or larger in order to keep consistency with the collider
constraints \cite{Salazar:2015gxa}, the constraints from the experimental
data on $K$, $D$ and $B$-meson mixings \cite{Huyen:2012uk} and $%
B_{s,d}\rightarrow \mu ^{+}\mu ^{-}$, $B_{d}\rightarrow K^{\ast }(K)\mu
^{+}\mu ^{-}$ decays \cite%
{CarcamoHernandez:2005ka,Martinez:2008jj,Buras:2013dea,Buras:2014yna,Buras:2012dp}

The $SU\left( 3\right) _{L}$ scalar triplets of our model can be expanded
around the mininum as follows: 
\begin{eqnarray}
\eta &=& 
\begin{pmatrix}
\frac{1}{\sqrt{2}}(v_{\eta }+\xi _{\eta }\pm i\zeta _{\eta }) \\ 
\eta _{2}^{-} \\ 
\eta _{3}^{0}%
\end{pmatrix}
,\hspace{0.5cm}\hspace{0.5cm}\rho = 
\begin{pmatrix}
\rho _{1}^{+} \\ 
\frac{1}{\sqrt{2}}(v_{\rho }+\xi _{\rho }\pm i\zeta _{\rho }) \\ 
\rho _{3}^{+}%
\end{pmatrix}
\\
\chi &=& 
\begin{pmatrix}
\chi _{1}^{0} \\ 
\chi _{2}^{-} \\ 
\frac{1}{\sqrt{2}}(v_{\chi }+\xi _{\chi }\pm i\zeta _{\chi })%
\end{pmatrix}
{.}\label{VEV}
\end{eqnarray}

The $SU(3)_{L}$ fermionic antitriplets and triplets are 
\begin{equation}
Q_{nL}= 
\begin{pmatrix}
D_{n} \\ 
-U_{n} \\ 
J_{n} \\ 
\end{pmatrix}
_{L},\hspace{0.5cm}Q_{3L}= 
\begin{pmatrix}
U_{3} \\ 
D_{3} \\ 
T \\ 
\end{pmatrix}
_{L},\hspace{0.5cm}L_{iL}= 
\begin{pmatrix}
\nu _{i} \\ 
l_{i} \\ 
\nu _{i}^{c} \\ 
\end{pmatrix}
_{L},\hspace{0.5cm}n=1,2,\hspace{0.1cm}i=1,2,3.
\end{equation}
where $l_{1,2,3}=e,\mu ,\tau $.

\begin{table}[tbp]
\begin{tabular}{|c|c|c|c|c|}
\hline
& $SU\left( 3\right) _{C}$ & $SU\left( 3\right) _{L}$ & $U\left( 1\right)
_{X}$ & $U\left( 1\right)$ \\ \hline
$\chi$ & $\mathbf{1}$ & $\mathbf{3}$ & $-\frac{1}{3}$ & $0$ \\ \hline
$\eta$ & $\mathbf{1}$ & $\mathbf{3}$ & $\mathbf{-}\frac{1}{3}$ & $4$ \\ 
\hline
$\rho$ & $\mathbf{1}$ & $\mathbf{3}$ & $\frac{2}{3}$ & $-2$ \\ \hline
$S^{-}$ & $\mathbf{1}$ & $\mathbf{1}$ & $-1$ & $\Antonio{2}$ \\ \hline
$\sigma$ & $\mathbf{1}$ & $\mathbf{1}$ & $0$ & $-2$ \\ \hline
\end{tabular}%
\caption{Scalar assigments under $SU\left( 3\right) _{C}\times SU\left(
3\right) _{L}\times U\left( 1\right) _{X}\times U(1)$.}
\label{scalars}
\end{table}

\begin{table}[tbp]
\begin{tabular}{|c|c|c|c|c|}
\hline
& $SU\left( 3\right) _{C}$ & $SU\left( 3\right) _{L}$ & $U\left( 1\right)
_{X}$ & $U\left( 1\right)$ \\ \hline
$Q_{nL}$ & $\mathbf{3}$ & $\mathbf{\overline{3}}$ & $0$ & $0$ \\ \hline
$Q_{3L}$ & $\mathbf{3}$ & $\mathbf{3}$ & $\frac{1}{3}$ & $0$ \\ \hline
$u_{iR}$ & $\mathbf{3}$ & $\mathbf{1}$ & $\frac{2}{3}$ & $-4$ \\ \hline
$d_{iR}$ & $\mathbf{3}$ & $\mathbf{1}$ & $-\frac{1}{3}$ & $2$ \\ \hline
$J_{1R}$ & $\mathbf{3}$ & $\mathbf{1}$ & $\frac{2}{3}$ & $0$ \\ \hline
$J_{nR}$ & $\mathbf{3}$ & $\mathbf{1}$ & $-\frac{1}{3}$ & $0$ \\ \hline
$T_{nL}$ & $\mathbf{3}$ & $\mathbf{1}$ & $\frac{2}{3}$ & $-2$ \\ \hline
$T_{nR}$ & $\mathbf{3}$ & $\mathbf{1}$ & $\frac{2}{3}$ & $-2$ \\ \hline
$B_{nL}$ & $\mathbf{3}$ & $\mathbf{1}$ & $-\frac{1}{3}$ & $4$ \\ \hline
$B_{nR}$ & $\mathbf{3}$ & $\mathbf{1}$ & $-\frac{1}{3}$ & $4$ \\ \hline
\end{tabular}%
\caption{Quark assigments under $SU\left( 3\right) _{C}\times SU\left(
3\right) _{L}\times U\left( 1\right) _{X}\times U(1)$. Here $i=1,2,3 $ and $%
n=1,2$.}
\label{quarks}
\end{table}

\begin{table}[tbp]
\begin{tabular}{|c|c|c|c|c|}
\hline
& $SU\left( 3\right) _{C}$ & $SU\left( 3\right) _{L}$ & $U\left( 1\right)
_{X}$ & $U\left( 1\right)$ \\ \hline
$L_{iL}$ & $\mathbf{1}$ & $\mathbf{3}$ & $-\frac{1}{3}$ & $1$ \\ \hline
\Antonio{$l_{iR}$} & $\mathbf{1}$ & $\mathbf{1}$ & $-1$ & $1$ \\ \hline
$N_{iR}$ & $\mathbf{1}$ & $\mathbf{1}$ & $0$ & $1$ \\ \hline
$E_{iL}$ & $\mathbf{1}$ & $\mathbf{1}$ & $-1$ & $3$ \\ \hline
$E_{iR}$ & $\mathbf{1}$ & $\mathbf{1}$ & $-1$ & $3$ \\ \hline
\end{tabular}%
\caption{Lepton assigments under $SU\left( 3\right) _{C}\times SU\left(
3\right) _{L}\times U\left( 1\right) _{X}\times U(1)$. Here $i=1,2,3 $.}
\label{leptons}
\end{table}
With the field assignment specified in Tables \ref{scalars}, \ref{quarks}
and \ref{leptons}, the following quark and lepton Yukawa terms arise: 
\begin{eqnarray}
-\mathcal{L}_{Y}^{\left( q\right) } &=&y_{1}^{\left( J\right) }\overline{Q}
_{3L}\chi J_{1R}+\sum_{n=1}^{2}y_{nm}^{\left( J\right) }\overline{Q}
_{nL}\chi ^{\ast }J_{mR}+\sum_{j=1}^{3}y_{3j}^{\left( u\right) }\overline{Q}
_{3L}\eta u_{jR}+\sum_{j=1}^{3}y_{3j}^{\left( d\right) }\overline{Q}
_{3L}\rho d_{jR}  \notag \\
&&\sum_{n=1}^{2}\sum_{m=1}^{2}y_{nm}^{\left( T\right) }\overline{Q}_{nL}\rho
^{\ast }T_{mR}+\sum_{n=1}^{2}\sum_{j=1}^{3}x_{nm}^{\left( u\right) } 
\overline{T}_{nL}\sigma ^{\ast }u_{jR}+\sum_{n=1}^{2}\sum_{m=1}^{2}\left(
M_{T}\right) _{nm}\overline{T}_{nL}T_{mR}  \notag \\
&&+\sum_{n=1}^{2}\sum_{m=1}^{2}y_{nm}^{\left( B\right) }\overline{Q}
_{nL}\eta ^{\ast }B_{mR}+\sum_{n=1}^{2}\sum_{j=1}^{3}x_{nj}^{\left( d\right)
}\overline{B}_{nL}\sigma ^{\ast }d_{jR}+\sum_{n=1}^{2}\sum_{m=1}^{2}\left(
M_{B}\right) _{nm}\overline{B}_{nL}B_{mR}+h.c  \label{lyq}
\end{eqnarray}
\begin{eqnarray}
-\mathcal{L}_{Y}^{\left( l\right) }
&=&\sum_{i=1}^{3}\sum_{j=1}^{3}y_{ij}^{\left( N\right) }\overline{L}
_{iL}\chi N_{jR}+\sum_{i=1}^{3}\sum_{j=1}^{3}x_{ij}^{\left( N\right) } 
\overline{E}_{iL}S^{-}N_{jR}+\sum_{i=1}^{3}\left( x_{\rho }^{\left( L\right)
}\right) _{ij}\varepsilon _{abc}\overline{L}_{iL}^{a}\left(
L_{jL}^{C}\right) ^{b}\left( \rho ^{\ast }\right) ^{c}  \notag \\
&&+\sum_{i=1}^{3}\sum_{j=1}^{3}y_{ij}^{\left( E\right) }\overline{L}
_{iL}\rho E_{jR}+\sum_{i=1}^{3}\sum_{j=1}^{3}x_{ij}^{\left( l\right) } 
\overline{E}_{iL}\sigma ^{\ast }l_{jR}+\sum_{i=1}^{3}\sum_{j=1}^{3}\left(
m_{E}\right) _{ij}\overline{E}_{iL}E_{jR}+h.c  \label{lyl}
\end{eqnarray}
In what follows we comment about some phenomenological aspects of our model
concerning LHC signals of non-SM fermions. As follows from the Yukawa terms
of Eq. (\ref{lyq}) and \ref{lyl}, the exotic fermions have mixing mass terms
with all the SM quarks, excepting the top and bottom quarks. Such mixing terms
allow that these exotic charged fermions can decay into any of the scalars
of the model and SM charged fermions. These exotic charged fermions can
decay into a SM charged fermion and a scalar. These heavy charged exotic
fermions can be pair produced at the LHC via gluon fusion (for the exotic
quarks only) and Drell Yan mechanism. Consequently, observing an excess of
events in the multijet and multilepton final state can be a signal of
support of this model at the LHC.

\section{Fermion mass matrices}
\label{fermionmasses}
From the quark Yukawa interactions in Eq. (\ref{lyq}), we find that the
up-type mass matrix in the basis $(\overline{u}_{1L},\overline{u}_{2L}, 
\overline{u}_{3L},\overline{J}_{1L},\overline{T}_{1L},\overline{T}_{2L})$
versus $(u_{1R},u_{2R},u_{3R},J_{1R},T_{1R},T_{2R})$ is given by: 
\begin{eqnarray}
M_{U} &=&\left( 
\begin{array}{ccc}
C_{U} & 0_{3\times 1} & A_{U} \\ 
0_{1\times 3} & m_{J_{1}} & 0_{1\times 2} \\ 
B_{U} & 0_{2\times 1} & M_{T}%
\end{array}
\right) ,\hspace{0.1cm}\hspace{0.1cm}\left( A_{U}\right) _{3j}=0,\hspace{
0.1cm}\hspace{0.1cm}\left( A_{U}\right) _{nm}=y_{nm}^{\left( T\right) }\frac{
v_{\rho }}{\sqrt{2}},\hspace{0.5cm}\left( B_{U}\right) _{nj}=x_{nj}^{\left(
u\right) }v_{\sigma },  \notag \\
C_{U} &=&\left( 
\begin{array}{ccc}
0 & 0 & 0 \\ 
0 & 0 & 0 \\ 
y_{31}^{\left( u\right) } & y_{32}^{\left( u\right) } & y_{33}^{\left(
u\right) }%
\end{array}
\right) \frac{v_{\eta }}{\sqrt{2}},\hspace{0.5cm}m_{J_{1}}=y_{1}^{\left(
J\right) }\frac{v_{\chi }}{\sqrt{2}},\hspace{0.5cm}j=1,2,3,  \label{MU}
\end{eqnarray}

whereas the down type quark mass matrix written in the basis $(\overline{d}
_{1L},\overline{d}_{2L},\overline{d}_{3L},\overline{J}_{1L},\overline{J}
_{2L},\overline{B}_{1L},\overline{B}_{2L})$-$%
(d_{1R},d_{2R},d_{3R},J_{1R},J_{2R},B_{1R},B_{2R})$ takes the form: 
\begin{eqnarray}
M_{D} &=&\left( 
\begin{array}{ccc}
C_{D} & 0_{3\times 2} & A_{D} \\ 
0_{2\times 3} & M_{J} & 0_{2\times 3} \\ 
B_{D} & 0_{3\times 2} & M_{B}%
\end{array}
\right) ,\hspace{0.1cm}\hspace{0.1cm}\left( A_{D}\right) _{3j}=0,\hspace{
0.1cm}\hspace{0.1cm}\left( A_{D}\right) _{nm}=y_{nm}^{\left( B\right) }\frac{
v_{\eta }}{\sqrt{2}},\hspace{0.1cm}\hspace{0.1cm}\left( B_{D}\right)
_{nj}=x_{nj}^{\left( d\right) }v_{\sigma },  \notag \\
C_{D} &=&\left( 
\begin{array}{ccc}
0 & 0 & 0 \\ 
0 & 0 & 0 \\ 
y_{31}^{\left( d\right) } & y_{32}^{\left( d\right) } & y_{33}^{\left(
d\right) }%
\end{array}
\right) \frac{v_{\rho }}{\sqrt{2}},\hspace{0.1cm}\hspace{0.1cm}\hspace{0.1cm}
\left( M_{J}\right) _{nm}=y_{nm}^{\left( J\right) }\frac{v_{\chi }}{\sqrt{2}}
,\hspace{0.5cm}n,m=1,2,\hspace{0.5cm}j=1,2,3.  \label{MD}
\end{eqnarray}

Furthermore, from the charged lepton Yukawa terms, we get that the mass
matrix for SM charged leptons in the basis $(\overline{l}_{1L},\overline{l}
_{2L},\overline{l}_{3L},\overline{E}_{1L},\overline{E}_{2L},\overline{E}
_{3L})$-$(l_{1R},l_{2R},l_{3R},E_{1R},E_{2R},E_{3R})$ is given by:

\begin{equation}
M_{E}=\left( 
\begin{array}{cc}
0_{3\times 3} & A_{E} \\ 
B_{E} & M_{E}%
\end{array}
\right) ,\hspace{0.1cm}\hspace{0.1cm}\hspace{0.1cm}\left( A_{E}\right)
_{ij}=y_{ij}^{\left( E\right) }\frac{v_{\rho }}{\sqrt{2}},\hspace{0.5cm}
\left( B_{E}\right) _{ij}=x_{ij}^{\left( l\right) }v_{\sigma },\hspace{0.5cm}
i,j=1,2,3.  \label{ME}
\end{equation}

The masses of the charged exotic vector like fermions are assumed to be much
larger than the $SU(3)_{L}\times U(1)_{X}$ symmetry breaking scale.
Consequently, the first and second generation of SM quarks as well as the SM
charged exotic leptons obtain their masses via Universal Seesaw mechanism,
while the bottom and top quark masses are generated from the renormalizable
Yukawa interactions involving the $SU\left( 3\right) _{L}$ scalar triplets $%
\rho $ and $\eta $, respectively. Furthemore, the exotic fermion masses are
generated at tree level. In view of the aforementioned considerations, we
find that the SM charged fermion mass matrices are given by:

\begin{eqnarray}
\widetilde{M}_{U} &=&C_{U}+A_{U}M_{\widetilde{T}}^{-1}B_{U}, \\
\widetilde{M}_{D} &=&C_{D}+A_{D}M_{B}^{-1}B_{D}, \\
\widetilde{M}_{E} &=&A_{E}M_{E}^{-1}B_{E}.
\end{eqnarray}

\begin{figure}[th]
\includegraphics[width=0.9\textwidth]{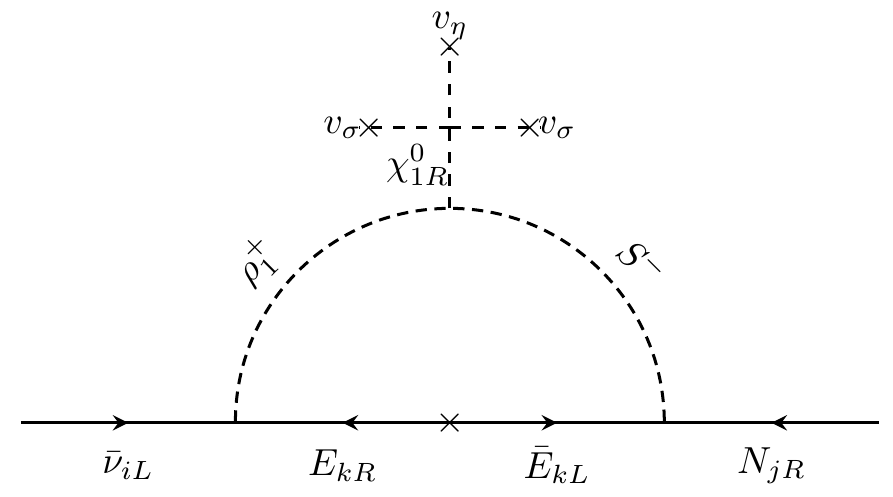} 
\caption{One-loop Feynman diagram contributing to the neutrino mass
submatrix $\protect\varepsilon$. Here $i,j,k,l=1,2,3$.}
\label{Neutrinoloopdiagram}
\end{figure}

The neutrino Yukawa interactions give rise to the following neutrino mass
terms:

\begin{equation}
-\mathcal{L}_{mass}^{\left( \nu \right) }=\frac{1}{2}\left( 
\begin{array}{ccc}
\overline{\nu _{L}^{C}} & \overline{\nu _{R}} & \overline{N_{R}}%
\end{array}
\right) M_{\nu }\left( 
\begin{array}{c}
\nu _{L} \\ 
\nu _{R}^{C} \\ 
N_{R}^{C}%
\end{array}
\right) +H.c,
\end{equation}

where the neutrino mass matrix reads:

\begin{equation}
M_{\nu }=\left( 
\begin{array}{ccc}
0_{3\times 3} & m_{\nu D} & \varepsilon \\ 
m_{\nu D}^{T} & 0_{3\times 3} & M \\ 
\varepsilon ^{T} & M^{T} & 0_{3\times 3}%
\end{array}
\right) ,  \label{Mnu}
\end{equation}

being the submatrices $M$ and $m_{\nu D}$ generated at tree level from the
first and second terms of the leptonic Yukawa interactions of Eq. (\ref{lyl}
), respectively, whereas the submatrix $\varepsilon $ is generated at one
loop level from the Feynman diagram of Figure \ref{Neutrinoloopdiagram}. The
aforementioned submatrices are given by:

\begin{eqnarray}
m_{\nu D} &=&\left[ \left( x_{\rho }^{\left( L\right) }\right) ^{\dagger
}-\left( x_{\rho }^{\left( L\right) }\right) ^{\ast }\right] \frac{v_{\rho } 
}{2\sqrt{2}},\hspace{0.1cm}\hspace{0.1cm}\hspace{0.1cm}M_{ij}=y_{ij}^{\left(
N\right) }\frac{v_{\chi }}{\sqrt{2}},\hspace{0.5cm}i,j=1,2,3, \\
\varepsilon _{ij} &=&\frac{\lambda _{S^{+}\rho ^{\dagger }\chi \sigma }}{
16\pi ^{2}}\sum_{k=1}^{3}\frac{y_{ik}^{\left( E\right) }x_{kj}^{\left(
N\right) }v_{\sigma }v_{\chi }}{m_{E_{k}}}C_{0}\left( \frac{m_{\rho
_{1}^{+}} }{m_{E_{k}}},\frac{m_{S^{-}}}{m_{E_{k}}}\right)
\end{eqnarray}

and the loop function has the form: 
\begin{equation}
C_{0}\left( \widehat{m}_{1},\widehat{m}_{2}\right) =\frac{\widehat{m}%
_{1}^{2} \widehat{m}_{2}^{2}\ln \left( \frac{\widehat{m}_{1}^{2}}{\widehat{m}%
_{2}^{2}} \right) -\widehat{m}_{1}^{2}\ln \widehat{m}_{1}^{2}+\widehat{m}%
_{2}^{2}\ln \widehat{m}_{2}^{2}}{\left( 1-\widehat{m}_{1}^{2}\right) \left(
1-\widehat{m} _{2}^{2}\right) \left( \widehat{m}_{1}^{2}-\widehat{m}%
_{2}^{2}\right) }.  \label{loopfunction}
\end{equation}

The light active neutrino masses arise from a linear seesaw mechanism and
the physical neutrino mass matrices are: 
\begin{eqnarray}
M_{\nu }^{\left( 1\right) } &=&-\left[ \varepsilon M^{-1}m_{\nu
D}^{T}+m_{\nu D}\left( M^{T}\right) ^{-1}\varepsilon ^{T}\right] , \\
M_{\nu }^{\left( 2\right) } &=&-\frac{1}{2}\left( M+M^{T}\right) -\frac{1}{2}
\left[ m_{\nu D}^{T}m_{\nu D}\left( M^{T}\right) ^{-1}+\left( M\right)
^{-1}m_{\nu D}^{T}m_{\nu D}\right] , \\
M_{\nu }^{\left( 3\right) } &=&\frac{1}{2}\left( M+M^{T}\right) +\frac{1}{2} %
\left[ m_{\nu D}^{T}m_{\nu D}\left( M^{T}\right) ^{-1}+\left( M\right)
^{-1}m_{\nu D}^{T}m_{\nu D}\right] ,
\end{eqnarray}
where $M_{\nu }^{\left( 1\right) }$ corresponds to the active neutrino mass
matrix whereas $M_{\nu }^{\left( 2\right) }$ and $M_{\nu }^{\left( 3\right)
} $ are the sterile neutrino mass matrices. The smallness of the light
active neutrino masses is attributed to the loop suppression as well as to
the small $\frac{\left( m_{\nu D}\right) _{ij}}{M_{ij}}\sim \frac{v_{\rho }}{
v_{\chi }}$ ratio. The physical neutrino spectrum is composed of 3 light
active neutrinos and 6 nearly degenerate sterile exotic pseudo-Dirac
neutrinos. The sterile neutrinos can be produced in pairs at the LHC, via
quark-antiquark annihilation mediated by a heavy $Z^\prime $ gauge boson.
They can decay into SM particles giving rise to a SM charged lepton and a $W$
gauge boson in the final state. Thus, observing an excess of events with
respect to the SM background in the opposite sign dileptons final states can
be a signal in support of this model at the LHC.

\section{The scalar potential}
\label{scalarpotential}
The most general scalar potential, with one neutral scalar singlet, one electrically charged scalar singlet, and three scalar triplets, is written as follows:
\begin{align}
V_{\left( \eta ,\rho ,\chi ,\sigma ,S\right) }& =\mu _{\chi }^{2}\left( \chi
^{\dagger }\chi \right) +\mu _{\eta }^{2}\left( \eta^{\dagger }\eta \right)
+\mu _{\rho }^{2}\left( \rho^{\dagger }\rho \right) +\mu _{\sigma}\left( \left(
\sigma ^{\dagger }\sigma \right) +h.c\right) +\mu _{S}^{2}S^{+}S^{-}+\lambda _{1}\left( \chi ^{\dagger }\chi \right) \left( \chi ^{\dagger
}\chi \right)   \notag \\
& +\lambda _{2}\left( \rho ^{\dagger }\rho \right) \left( \rho ^{\dagger
}\rho \right) +\lambda _{3}\left( \eta ^{\dagger }\eta \right) \left( \eta
^{\dagger }\eta \right) +\lambda _{4}\left( \sigma ^{\dagger }\sigma \right)
^{2}+\lambda _{5}\left( \chi ^{\dagger }\chi \right) \left( \sigma ^{\dagger
}\sigma \right) +\lambda _{6}\left( \eta ^{\dagger }\eta \right) \left(
\sigma ^{\dagger }\sigma \right)   \notag \\
& +\lambda _{7}\left( \rho ^{\dagger }\rho \right) \left( \sigma ^{\dagger
}\sigma \right) +\lambda _{8}\left( \chi ^{\dagger }\chi \right) \left( \rho
^{\dagger }\rho \right) +\lambda _{9}\left( \chi ^{\dagger }\chi \right)
\left( \eta ^{\dagger }\eta \right) +\lambda _{10}\left( \rho ^{\dagger
}\rho \right) \left( \eta ^{\dagger }\eta \right) +\lambda _{11}\left( \chi ^{\dagger }\eta \right) \left( \eta ^{\dagger
}\chi \right)   \label{VS} \\
& +\lambda _{12}\left( \chi ^{\dagger }\rho \right) \left( \rho
^{\dagger }\chi \right) +\lambda _{13}\left( \rho ^{\dagger }\eta \right)
\left( \eta ^{\dagger }\rho \right) -\frac{f_{q}}{\sqrt[2]{2}}\left( \sigma
\chi _{i}\eta _{j}\rho _{k}\varepsilon ^{ijk}+\text{ H.c. }\right)  +\frac{f_{s}}{\sqrt[2]{2}}\left( \left( S^{+}S^{-}\right) \left( \sigma
^{\dagger }\sigma \right) +\text{ H.c. }\right) \notag
\\
&  +\frac{f}{\sqrt[2]{2}}%
\left( \Antonio{(\chi ^{\dagger }\eta )\sigma ^2} +\text{
H.c. }\right) +
\Antonio{\frac{A}{\sqrt[2]{2}}\left( \chi ^{\dagger }S^{-}\rho +\text{ H.c. }\right)}  \notag \\
& +\lambda_{14}\left( S^{+}S^{-}\right) ^{2}+\lambda_{15}\left( \chi
^{\dagger }\chi \right) \left( S^{+}S^{-}\right) +\lambda_{16}\left(
\eta ^{\dagger }\eta \right) \left( S^{+}S^{-}\right) +\lambda_{17}\left( \rho ^{\dagger }\rho \right) \left( S^{+}S^{-}\right) .  \notag
\end{align}

Where $\mu _{a}$ $(a=\chi ,\mu ,\rho ,\sigma )$ are dimensionful parameters
(mass matrices), coefficiens $\lambda_{i}$ $(i=1,...,17)$, \Yoxara{ $f$ and $f_{j}$ $%
(j=q,s,)$ are quartic scalar couplings, and $A$ is the trilinear coupling. } 
Besides, charged scalar singlet $%
S^{-}$ is not allowed to have VEV, and we take on that the neutral scalars $\eta
_{3}^{0}$ and $\chi _{1}^{0}$ do not develop VEVs. Therefore, we can obtain $%
\mu _{a}$ from the scalar potential Eq. (\ref{VS}) expanding the scalars
around the VEVs as in Eq. (\ref{VEV}), and using the minimum conditions,

\begin{align}
	& \mu _{\chi }^{2}=-\lambda_{1}v_{\chi }^{2}-\frac{\lambda_{5}}{2}v_{\sigma
	}^{2}-\frac{\lambda_{8}}{2}v_{\rho }^{2}-\frac{\lambda_{9}}{2}v_{\eta
	}^{2}+f_{q}\frac{v_{\eta }v_{\rho }v_{\sigma }}{2\sqrt{2}v_{\chi }} \notag
	\\
	& \mu _{\eta }^{2}=-\lambda_{3}v_{\eta }^{2}-\frac{\lambda_{6}}{2}v_{\sigma
	}^{2}-\frac{\lambda_{9}}{2}v_{\chi }^{2}-\frac{\lambda_{10}}{2}v_{\rho
	}^{2}+f_{q}\frac{v_{\chi }v_{\rho }v_{\sigma }}{2\sqrt{2}v_{\eta }} \notag
	\\
	& \mu _{\rho }^{2}= -\lambda_{2}v_{\rho }^{2}-\frac{\lambda_{7}}{2}v_{\sigma
	}^{2}-\frac{\lambda_{8}}{2}v_{\chi }^{2}+\frac{\lambda_{10}}{2}v_{\eta
	}^{2}+f_{q}\frac{v_{\chi }v_{\eta }v_{\sigma }}{2\sqrt{2}v_{\rho }}  \notag
	\\
	& \mu _{\sigma }^{2}=-\frac{\lambda_{4}}{2}v_{\sigma }^{2}-\frac{\lambda_{5}%
	}{4}v_{\chi }^{2}-\frac{\lambda_{6}}{4}v_{\eta }^{2}-\frac{\lambda_{7}}{4}%
	v_{\rho }^{2}+f_{q}\frac{v_{\eta }v_{\rho }v_{\chi }}{4\sqrt{2}v_{\sigma }}%
	.  \label{MUs}
\end{align}%

Substituting Eq.  (\ref{MUs}), and Eq. (\ref{VEV}) in the scalar potential Eq. %
(\ref{VS}), we can determinate three mass squared matrices for CP odd, CP
even, and complex neutral scalars in the pseudoscalar basis $\left( \zeta
_{\eta },\zeta _{\rho },\zeta _{\chi },\sigma _{I}\right) $, scalar basis 
$\left( \xi _{\eta },\xi _{\rho },\xi _{\chi },\sigma_{R}\right) $ and
complex scalar basis $\left( \eta _{3}^{0},\chi _{1}^{0}\right) $\
respectively. Also, we can obtain a mass squared matrix for charged scalars
in the basis $\left( \eta _{2}^{+}\rho _{1}^{+},\rho _{3}^{+},\chi
_{2}^{+},S^{+}\right) $. In this way, we find the physical mass eigenvalues of these square matrices. Thus, for the pseudoscalar basis we have the squared masses for the CP odd $A_{1}$ physical scalar field and the Goldstone bosons $G_{1}^{0}$, $G_{2}^{0}$ and $G_{3}^{0}$ as following,

\begin{align}
& M_{A_{1}}^{2}=\frac{f_{q}(v_{\chi }^{2}v_{\sigma }^{2}v_{\rho
}^{2}+v_{\eta }^{2}(v_{\chi }^{2}v_{\sigma }^{2}+(v_{\chi }^{2}+v_{\sigma
}^{2})v_{\rho }^{2}))}{2\sqrt{2}v_{\chi }v_{\eta }v_{\rho }v_{\sigma }}\notag \\
& M_{G_{1}^{0}}^{2}=M_{G_{2}^{0}}^{2}=M_{G_{3}^{0}}^{2}=0,
\end{align}
being $A_{1}$ a heavy pseudoscalar. For the complex scalar basis, we can obtain the following squared mass,
\begin{align}
&M_{\phi }^{2}=\frac{v_{\chi }(\sqrt{2}f_{q}v_{\sigma }v_{\rho }+2\lambda_{11}v_{\chi }v_{\eta })}{4v_{\eta }}. \notag \\
&M_{G_{\phi} }^{2}=0,
\end{align}
where $G_{\phi}$ is the Goldstone boson corresponding to the imaginary part of the gauge singlet scalar $\varphi$. Such Goldstone boson associated with the spontaneous breaking is the global $U(1)$ symmetry is the Majoron which is harmless since it is a gauge singlet. It is worth mentioning that the imaginary part of $\varphi$ can acquire a mass by including a soft breaking term $\mu^2_{sb}\left(\varphi^2+h.c\right)$ in the scalar potential. Analyzing the squared mass matrix for the CP even scalar fields,

\begin{eqnarray}
\label{eq:CP-even-1}
 M^2_{CP} =\left(
\begin{array}{cccc}
	\frac{\sqrt{2} f_{q} v_{\sigma}  v_{\rho} v_{\chi}+8 \lambda _3 v_{\eta}^3}{4 v_{\eta}} & \lambda _{10} v_{\eta} v_{\rho}-\frac{f_{q} v_{\sigma}  v_{\chi}}{2
		\sqrt{2}} & \lambda _9 v_{\eta} v_{\chi}-\frac{f_{q} v_{\sigma} v_{\rho}}{2 \sqrt{2}} & \lambda _6 v_{\sigma} v_{\eta}-\frac{f_{q} v_{\rho}
		v_{\chi}}{2 \sqrt{2}} \\
	\lambda _{10} v_{\eta} v_{\rho}-\frac{f_{q} v_{\sigma}  v_{\chi}}{2 \sqrt{2}} & \frac{\sqrt{2} f_{q} v_{\sigma}  v_{\eta} v_{\chi}+8 \lambda _2
		v_{\rho}^3}{4 v_{\rho}} & \lambda _8 v_{\rho} v_{\chi}-\frac{f_{q} v_{\sigma} v_{\eta}}{2 \sqrt{2}} & \lambda _7 v_{\sigma}  v_{\rho}-\frac{f_{q} v_{\eta}
		v_{\chi}}{2 \sqrt{2}} \\
	\lambda _9 v_{\eta} v_{\chi}-\frac{f_{q} v_{\sigma} v_{\rho}}{2 \sqrt{2}} & \lambda _8 v_{\rho} v_{\chi}-\frac{f_{q} v_{\sigma}  v_{\eta}}{2 \sqrt{2}} &
	\frac{\sqrt{2} f_{q} v_{\sigma}  v_{\eta} v_{\rho}+8 \lambda _1 v_{\chi}^3}{4 v_{\chi}} & \lambda _5 v_{\sigma}  v_{\chi}-\frac{f_{q} v_{\eta} v_{\rho}}{2
		\sqrt{2}} \\
	\lambda _6 v_{\sigma} v_{\eta}-\frac{f_{q} v_{\rho} v_{\chi}}{2 \sqrt{2}} & \lambda _7 v_{\sigma} v_{\rho}-\frac{f_{q} v_{\eta} v_{\chi}}{2 \sqrt{2}} &
	\lambda _5 v_{\sigma}  v_{\chi}-\frac{f_{q} v_{\eta} v_{\rho}}{2 \sqrt{2}} & \frac{\sqrt{2} f_{q} v_{\eta} v_{\rho} v_{\chi}+8 \lambda _4 v_{\sigma}
		^3}{4 v_{\sigma}} \\
\end{array}
\right),
\end{eqnarray}
we can find physical scalars   $h_{1}^{0}$\ , $H_{1}$\,$ H_{2}$ and their masses $M_{h_{1}^{0}}^{2}$, $M_{H_{1}}^{2}$ and $ M_{H_{2}}^{2}$. 
In order to simplify the computations we consider a benchmark point: \Yoxara{ $\lambda_{6} = \lambda_{7} \left(\frac{v_{\rho}}{v_{\eta}}\right)^{2}$, $f_{q} = \frac{4\sqrt{2}v_{\sigma }^{3}}{v_{\chi } v_{\rho} v_{\eta }}\lambda_{4}$, $\lambda_{8} = \lambda_{9} \left(\frac{v_{\eta}}{v_{\rho}}\right)^{2}$, $\lambda_{5} = \lambda_{9} \left(\frac{v_{\eta}}{v_{\sigma}}\right)^{2}$, $ \lambda_{7}=\lambda_{9} \left(\frac{v_{\chi}}{v_{\sigma}}\right)^{2} \left(\frac{v_{\eta}}{v_{\rho}}\right)^{2}$, $\lambda_{10} = 2\lambda_{3}$, $\lambda_{2}=\lambda_{3}$ and $\lambda_{9} = \frac{2v_{\sigma}^{4}}{v_{\chi}^{2} v_{\eta}^{2}}\lambda_{4}$.}
In this case 
we find from (\ref{eq:CP-even-1}) for the squared masses for the CP even Higgs bosons, $M_{h_{1}^{0}}^{2}$, $ M_{H_{1}}^{2}$, $ M_{H_{2}}^{2}$ and the Goldstone boson $ M_{G_{H}}^{2}$ the following expressions
\begin{align}
&M_{h_{1}^{0}}^{2}= 2v^{2} \lambda_{3},
\notag\\
&  M_{H_{1}}^{2}= 2 v_{\chi }^{2} \lambda_{1} +2 \frac{v_{\sigma }^{4}}{v_{\chi }^{2}} \lambda_{4}
\notag\\
& M_{H_{2}}^{2} = 2 v_{\sigma }^{4} \lambda_{4} \left(\frac{1}{v_{\eta}^{2}}+\frac{1}{v_{\rho}^{2}}\right), 
\quad M_{H_{3}}^{2}=4\lambda_{4}v_{\sigma}^{2}
\end{align}
Hence, the scalar denoted as $h^{0}_{1}$ might be linked to the SM Higgs boson; the scalar $H_{1}$ would correspond to the physical scalar field associated with the spontaneous symmetry breaking produced through the $SU(3)_{L}$ scalar triplet $\chi$, and the \Yoxara{scalars $H_{2}$ and $H_{3}$ could be identified with the CP even part of the $\sigma$ field}. Finally, the squared masses for the charged scalar fields are,
\begin{align}
& M_{h_{1}^{\pm}}^{2}=\frac{(v_{\chi }^{2}+v_{\rho }^{2})(\sqrt{2}%
f_{q}v_{\sigma }v_{\eta}+2\lambda_{12}v_{\chi }v_{\rho})}{4v_{\chi
}v_{\rho}}\notag \\
& M_{h_{2}^{\pm}}^{2}=\frac{(v^{2}(\sqrt{2}%
f_{q}v_{\sigma }v_{\chi}+2\lambda_{13}v_{\eta }v_{\rho})}{4v_{\eta
}v_{\rho }},\notag\\ 
& M_{h_{3}^{\pm}}^{2}=\frac{1}{2}\left(\sqrt{2}f_{s}v_{\sigma }^{2}+\lambda_{15}v_{\chi }^{2}+\lambda_{16} v_{\eta}^{2}+\lambda_{17} v_{\rho}^{2} + 2\mu_{S}^{2}\right),\notag\\ 
&M_{G_{1}^{\pm}}  = M_{G_{2}^{\pm}}=0,
\end{align}
with $G_{1}^{\pm}$, $G_{1}^{\pm}$ as Goldstone bosons, and taking into account $f=0$. 
\newpage

\section{Muon anomalous magnetic moment}
\label{gminus2}
\begin{figure}[!h]
\includegraphics[width=0.9\textwidth]{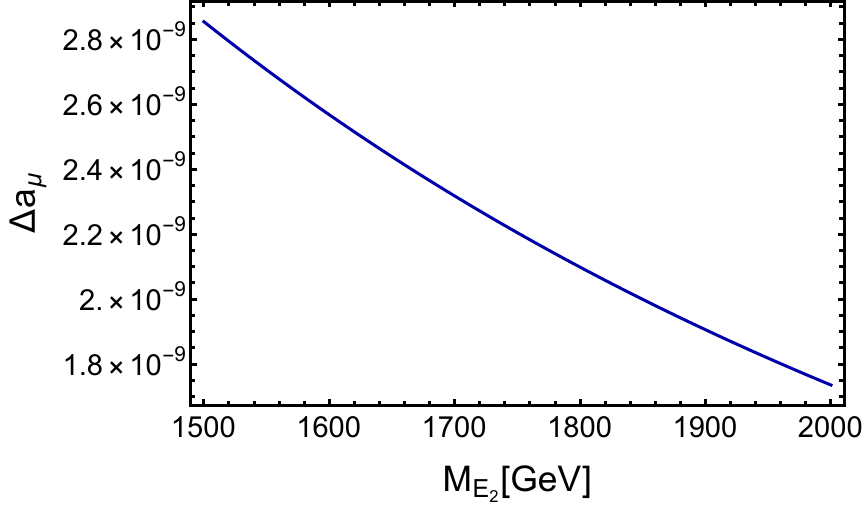}
\caption{Muon anomalous magnetic moment as a function of the charged exotic
lepton mass $M_{E_2}$.}
\label{gminus2muonvsmE}
\end{figure}

In this section we will analyze the phenomenological consequences of our
model in the muon anomalous magnetic moment. The dominant contribution to
the muon anomalous magnetic moment corresponds to the one-loop diagram
involving the exchange of electrically neutral CP even and CP odd scalars
and charged exotic lepton $E_{2}$ running in the internal lines of the loop. 
To simplify our analysis we can consider a simplified benchmark scenario
close to the decoupling limit where $\xi _{\rho }$ ($\zeta _{\rho }$) and $%
\sigma _{R}$ ($\sigma _{I}$) are mainly composed of two orthogonal
combinations involving two heavy CP even (odd) $H_{1}$ ($A_{1}$), $H_{2}$ ($%
A_{2}$) physical scalar fields. Furthermore, we work on the basis where the
SM charged lepton mass matrix is diagonal, which implies that the charged
exotic leptons $E_{i}$ ($i=1,2,3$) are physical fields \Antonio{and the submatrices $A_E$ and $B_E$ are diagonal.} Consequently, the muon anomalous magnetic moment is given by: 
\begin{equation}
\Delta a_{\mu }\simeq \frac{\Antonio{y_{22}^{\left( E\right)}x_{22}^{\left( E\right)}}m_{\mu }^{2}}{8\pi ^{2}}\left[
I_{S}\left( m_{E_{2}},m_{H_{1}}\right) -I_{S}\left(
m_{E_{2}},m_{H_{2}}\right) +I_{P}\left( m_{E_{2}},m_{A_{1}}\right)
-I_{P}\left( m_{E_{2}},m_{A_{2}}\right) \right] \sin \theta \cos \theta ,
\end{equation}
where $H_{1}\simeq \cos \theta _{S}\sigma _{R}+\sin \theta _{S}\xi _{\rho }$
, $H_{2}\simeq -\sin \theta _{S}\sigma _{R}+\cos \theta _{S}\xi _{\rho }$, $%
A_{1}\simeq \cos \theta _{P}\sigma _{I}+\sin \theta _{P}\zeta _{\rho }$, $%
A_{2}\simeq -\sin \theta _{P}\sigma _{I}+\cos \theta _{P}\zeta _{\rho }$ and
for the sake of simplicity we have set $\theta _{S}=\theta _{P}$ and $%
m_{E_{2}}$ is the mass of the charged exotic lepton $E_{2}$. Furthermore,
loop $I_{S\left( P\right) }\left( m_{E},m\right) $ function has the form 
\cite%
{Diaz:2002uk,Jegerlehner:2009ry,Kelso:2014qka,Lindner:2016bgg,Kowalska:2017iqv}
: 
\begin{equation}
I_{S\left( P\right) }\left( m_{E},m_{S,P}\right) =\int_{0}^{1}\frac{
x^{2}\left( 1-x\pm \frac{m_{E}}{m_{\mu }}\right) }{m_{\mu }^{2}x^{2}+\left(
m_{E}^{2}-m_{\mu }^{2}\right) x+m_{S,P}^{2}\left( 1-x\right) }dx
\end{equation}
Considering that the muon anomalous magnetic moment is constrained to be in
the range \cite{Hagiwara:2011af,Davier:2019can,Nomura:2018lsx,Nomura:2018vfz,Arbelaez:2020rbq,Yin:2021yqy,CarcamoHernandez:2021qhf,Abi:2021gix}: 
\begin{equation}
\left( \Delta a_{\mu }\right) _{\exp }=\left( 2.51\pm 0.59\right) \times
10^{-9}.
\end{equation}
We display in Figure \ref{gminus2muonvsmE} the muon anomalous magnetic
moment as a function of the charged exotic lepton mass $M_{E_2}$. In our
numerical analysis we have considered a benchmark scenario where we have
fixed $\theta =\frac{\pi }{4}$, $M_{A_1}=M_{H_1}=0.5$ TeV, $M_{A_2}=0.6$ TeV
and $M_{H_2}=2$ TeV. The mass of the charged exotic lepton $E_2$ has been
varied in the range $1.5$ TeV$\leqslant M_{E_2}\leqslant$ $2$ TeV. Figure \ref%
{gminus2muonvsmE} shows that the muon anomalous magnetic moment decreases
when the charged exotic lepton mass is increased. We find that our model can
successfully accommodate the experimental value of $\Delta a_{\mu}$. 

\newpage

\section{Conclusions}
It has been shown  that the most popular 3-3-1 models, such as 
minimal 3-3-1 are not capable of explaining g-2. They need to be extended. Moreover, the simplest versions of the most popular 3-3-1 models fail in generating two mass differences between the active neutrinos and be consistent with observation, unless a scalar sextet is added to the models. Motivated by these two facts, we proposed a new 3-3-1 model where neutrino masses are generated via one-loop linear seesaw and g-2 is accommodated in agreement with existing bounds. Our work, stands a plausible alternative to the well known 3-3-1 models in the literature, especially if the g-2 anomaly reaches $5\sigma$ of statistical significance with a larger data sample from the Muon g-2 experiment.
\label{conclusions}

\section*{Acknowledgements}
The authors thank Alfonso Zerwekh for discussions. AECH has received funding from ANID-Chile FONDECYT 1210378, ANID-Chile PIA/APOYO AFB180002 and Milenio-ANID-ICN2019\_044. SK acknowledges support from CONICYT-Chile Fondecyt No. 1190845, ANID-Chile PIA/APOYO AFB180002 and Milenio-ANID-ICN2019\_044. YSV acknowledge support from CAPES under grant number 88882.375870/2019-01. FSQ thanks CNPq grants 303817/2018-6 and 421952/2018-0, and Sao Paulo Research Foundation (FAPESP) through ICTP-SAIFR FAPESP grant 2016/01343-7,2021/01089-1, and grants 2015/158971 and Milenio-ANID-ICN2019\_044, for the financial support. FSQ thanks Universidad Frederico Santa Maria, Universidad Andres Bello and UFRGS for the hospitality where part of this was partly done. This work was supported by the Serrapilheira Institute (grant number Serra-1912-31613). We thank the High Performance Computing Center (NPAD) at UFRN for providing computational resources.\vspace{0.4cm}

\centerline{\bf{REFERENCES}}\vspace{-0.4cm}
\bibliographystyle{utphys}
\bibliography{Biblio331LSMarch1nd2022}
\end{document}